\documentclass[journal=nalefd,manuscript=article]{achemso}

\usepackage{graphicx}
\usepackage{dcolumn}
\usepackage[version=4]{mhchem}
\usepackage{bm}
\usepackage{xcolor}
\usepackage{soul}
\usepackage{sidenotes}

\author{S.~Kovalev}
\email{s.kovalev@hzdr.de}
\affiliation{Helmholtz Zentrum Dresden Rossendorf, Germany}
\author{K.-J.~Tielrooij}
\affiliation{Catalan Institute of Nanoscience and Nanotechnology (ICN2), BIST and CSIC, Campus UAB, Bellaterra (Barcelona), 08193, Spain}
\author{J.-C.~Deinert}
\author{I.~Ilyakov}
\author{N.~Awari}
\author{M.~Chen}
\author{A.~Ponomaryov}
\author{M.~Bawatna}
\affiliation{Helmholtz Zentrum Dresden Rossendorf, Germany}
\author{T.V.A.G. de ~Oliveira}
\affiliation{Technische Universit\"at Dresden, Germany}
\alsoaffiliation{ct.qmat - Cluster of Excellence on Complexity and Topology in Quantum Matter}
\author{L.M.~Eng}
\affiliation{Technische Universit\"at Dresden, Germany}
\alsoaffiliation{ct.qmat - Cluster of Excellence on Complexity and Topology in Quantum Matter}
\author{K.A.~Kuznetsov}
\email{kirill-spdc@yandex.ru}
\author{D.A.~Safronenkov}
\author{G.Kh.~Kitaeva}
\affiliation{Lomonosov Moscow State University, Russia}
\author{P.I.~Kuznetsov}
\affiliation{Kotelnikov IRE RAS, Russia}
\author{H.A.~Hafez}
\author{D.~Turchinovich}
\affiliation{Fakult\"at fur Physik, Universit\"at Bielefeld, Germany}
\author{M.~Gensch}
\affiliation{Helmholtz Zentrum Dresden Rossendorf, Germany}
\alsoaffiliation{DLR - Institute of Optical Sensor Systems, Germany}
\alsoaffiliation{Technische Universit\"at Berlin, Germany}

\title{Terahertz signatures of ultrafast Dirac fermion relaxation at the surface of topological insulators at room temperature}

\begin{document}


\newpage
\begin{abstract}
\textbf{Topologically-protected surface states present rich physics and promising spintronic, optoelectronic and photonic applications that require a proper understanding of their ultrafast carrier dynamics. Here, we investigate these dynamics in topological insulators (TIs) of the bismuth and antimony chalcogenide family, where we isolate the response of Dirac fermions at the surface from the response of bulk carriers by combining photoexcitation with below-bandgap terahertz (THz) photons with TI samples with varying Fermi level, including one sample with the Fermi level located within the bandgap. We identify distinctly faster relaxation of charge carriers in the topologically-protected Dirac surface states (few hundred femtoseconds), compared to bulk carriers (few picoseconds). In agreement with such fast cooling dynamics, we observe THz harmonic generation without any saturation effects for increasing incident fields, unlike graphene which exhibits strong saturation. This opens up promising avenues for increased THz nonlinear conversion efficiencies, and high-bandwidth optoelectronic and spintronic information and communication applications. }
\end{abstract}

Nowadays, due to their unique transport properties, topological insulators (TIs) attract great attention \cite{Fu2007, Burkov2010}. Due to the symmetry-protected Dirac fermions with non-trivial topology on their surface, they are highly prospective as functional materials in future electronic and spintronic devices \cite{Pesin2012}.
Moreover, various optoelectronic and thermoelectric applications can benefit from the combination of protected charge transport in surface states, and large Seebeck coefficients. For many of these applications it is crucial to understand the characteristic timescales of the relaxation dynamics of excited carriers, and in particular determine if these dynamics are different for topological surface states compared to the bulk.
\\

The carrier dynamics in TIs have been addressed using various pump-probe techniques \cite{Chen2010, Hsieh2011, Sobota2012, Wang2012, Glinka2013, Luo2013, Onishi2015, Luo2019}. However, despite these experimental efforts, the exact mechanism and timescale of carrier relaxation of the Dirac fermions in surface states of TIs are still under debate. The reason for this is that it is challenging to experimentally disentangle excitation and relaxation channels of surface and bulk states. This is caused by the relatively small bandgap of TIs of the bismuth and antimony chalcogenide family, which is a few hundred meV \cite{Arakane2012}, meaning that optical excitation with near-infrared and visible light leads to interband transitions between the bulk bands. Thus, in order to optically address the surface states, energetically located within the bandgap, photons in the mid-infrared or terahertz regimes are required. Furthermore, the Fermi energy is typically located either in the valence or the conduction band for binary TI compounds, e.g.\ \ce{Bi_2Se_3} and \ce{Bi_2Te_3}, such that bulk states are already populated. Whereas separating surface dynamics from bulk dynamics has recently been achieved at the cryogenic temperature of 5~K \cite{Luo2019}, this is not the case at more technologically relevant temperatures. The main reason for this is the occurrence of phonon-assisted surface-to-bulk scattering above the Debye temperature ($\sim$180~K for \ce{Bi_2Se_3}) \cite{Wang2012}. 
\\

Here, we overcome these challenges by combining optical excitation with low photon energies and a TI sample with the Fermi energy located inside the bandgap. As a result, we are able to isolate the response of Dirac electrons in the surface states, without the contribution of bulk states. In particular, we use THz pulses with photon energies below 4 meV, and verify whether the observed carrier dynamics originate from surface states (SS), bulk conduction states (BCS) or bulk valence states (BVS), using three different TI samples with the Fermi energy in the valence band (\ce{Bi_2Te_3}), in the conduction band (\ce{Bi_2Se_3}), and inside the bandgap (\ce{Bi_{1.4}Sb_{0.6}Te_{1.51}Se_{1.49}}, i.e.\ \ce{BSTS}). THz excitation of bulk-conducting \ce{Bi_2Se_3} and \ce{Bi_2Te_3} corresponds to intraband excitation of carriers in the bulk and at the surface simultaneously, such that surface and bulk relaxation dynamics cannot be disentangled in a straightforward manner. However, the BCS of the \ce{BSTS} are empty while the BVS are completely filled, so that intraband transitions in the bulk of \ce{BSTS} are Pauli blocked. As a result, in the \ce{BSTS} sample THz excitation is dominated by the intraband excitation of surface states only. Using two THz-based nonlinear techniques, namely THz-pump optical-probe (TPOP) and THz high-harmonic generation (THz HHG), we find that relaxation of Dirac fermions at the surface of TIs occurs on a timescale of a few hundred femtoseconds, which is significantly faster than the relaxation of bulk carriers.
\\


Our three different samples, rhombohedral \ce{Bi_2Se_3}, \ce{Bi_2Te_3} and \ce{Bi_{1.4}Sb_{0.6}Te_{1.51}Se_{1.49}} (\ce{BSTS}) films were grown on (0001) \ce{Al_2O_3} substrates. Their film thicknesses are 30 nm, 490 nm and 375 nm, respectively. 
The growth method is described in detail in Ref.\ \cite{Kuznetsov2018}. 
It was shown in \cite{Ren2011} that for topological insulators of the \ce{Bi_{2-x}Sb_xTe_{3-y}Se_y} quaternary compounds, there is an optimal curve (Ren's curve) in the composition diagram $y(x)$. Near this curve, the properties of electronic edge states are most pronounced due to the fact that acceptors and donors mutually cancel each other. Compositions below Ren's curve, as a rule of thumb, have predominant hole conductivity, while a composition above Ren's curve exhibits electronic conductivity. Using THz transmission measurements, we estimate the Fermi level of \ce{Bi_2Se_3} to be located 26 meV above the bottom of the conduction band, and for \ce{Bi_2Te_3} we find it 30 meV below the top of the valence band. In case of \ce{BSTS}, we find that the Fermi level is located inside the bandgap, being 110 meV below the conduction band, which is in agreement with earlier works \cite{Ren2011, Taskin2011, Arakane2012, Matsushita2017} (see the Supplementary information).
\\


We first measure the relaxation dynamics of these samples using transient reflectivity spectroscopy. We use laser pulses with 800 nm central wavelength and 100 fs pulse duration for probing, while single cycle THz pulses serve as a pump in THz-pump optical-probe (TPOP) experiments. As a reference, we perform optical-pump optical-probe (OPOP) measurements, where 800 nm pulses also serve as pump, instead of THz light. The experimental setups used for TPOP and OPOP measurements are described in the Supplementary information. Ultrafast pump-probe reflectivity measurements in the near- and mid-infrared ranges have been used earlier to study the dynamics of charge carriers \cite{Chen2010, Glinka2013, Luo2013, Onishi2015} and phonons \cite{Wu2008, Hu2018, Melnikov2018} in topological insulators. In these experiments, however, the pump pulses had photon energies above or close to the energy of the bandgap, such that the dynamics of bulk and surface carriers both contribute to the observed signals. In a few pump-probe studies, THz light was used as the pump, in order to study carriers \cite{Luo2019}, and phonons \cite{Melnikov2018}. However, in both cases bulk-metallic TIs were used. Therefore, the carrier relaxation dynamics observed in all of these studies always contained contributions from both the surface and bulk electronic systems.
\\

\begin{figure}[h]
\includegraphics{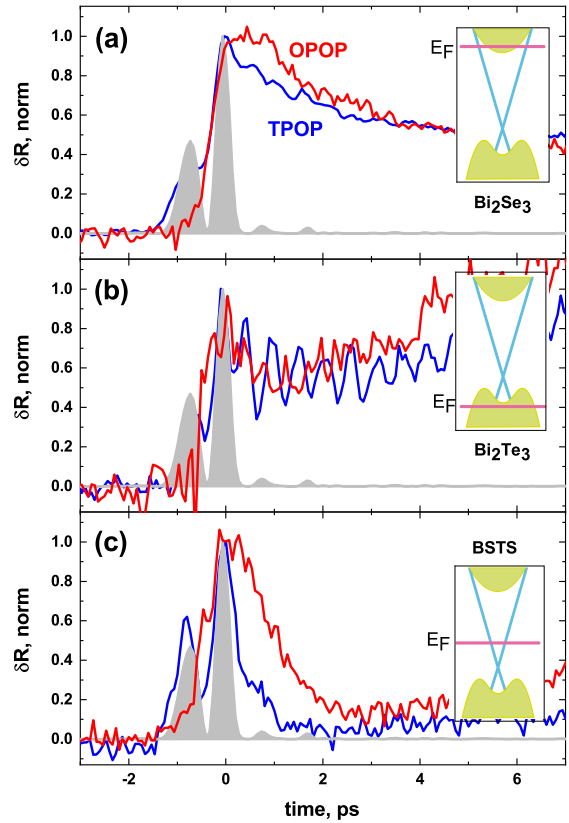}
\caption{\label{fig:TI} THz-pump optical-probe (TPOP) reflectivity measurements in \ce{Bi_2Se_3} (a), \ce{Bi_2Te_3} (b) and \ce{BSTS} (c) (blue line). The intensity transient of the driving THz field is shown in all panels (gray shaded line). The red line corresponds to the dynamics when using optical excitation with 1.5 eV photon energy (OPOP). The band diagram with the Fermi level of each sample, obtained from THz transmission experiments, is schematically shown in the insets.}
\end{figure} 

In Fig.~\ref{fig:TI} we show normalized optical-probe reflectivity changes for the three topological insulators investigated in our work after excitation by broadband THz radiation with 400 kV/cm peak field strength. The observed dynamics are significantly different for each of the three samples, and can be decomposed into several components: ultrafast build-up, bi-exponential decay and a Raman-excited coherent phonon response. The fast rise time of the THz response corresponds to the ultrafast thermalization of excited carriers through electron-electron scattering. As measured with optical techniques and time- and angle-resolved photoemission spectroscopy (tr-ARPES), the electron-electron scattering time constant has been determined to be lower than 200 fs for \ce{Bi_2Se_3} \cite{Wang2012}. The relaxation dynamics of \ce{Bi_2Se_3} (Fig.~\ref{fig:TI}~a, TPOP) can be decomposed into two different exponential decays.
The faster decay rate varies from 1.45 ps to 1.9 ps under excitation with 280 kV/cm and 400 kV/cm fields (see the Supplementary information) and can be attributed to electron-phonon scattering \cite{Luo2019, Wang2012, Hsieh2011, Glinka2013}. The slow decay channel shows a time constant of about 190 ps (see the Supplementary information). The presence of a slow relaxation upon sub-bandgap excitation in \ce{Bi_2Se_3} was observed previously using tr-ARPES \cite{Wang2013} and was attributed to thermal excitation of the carrier population. As mentioned before, due to the coexistence of partially filled BCS and gapless Dirac SS \cite{Xia2009}, it is not possible to separate relaxation processes occurring in bulk and surface states based on these measurements.
\\
 
Measurements performed on the \ce{Bi_2Te_3} sample are presented in Fig.~\ref{fig:TI}b. Here, we observe similar dynamics as for \ce{Bi_2Se_3}: ultrafast carrier build-up and a bi-exponential decay. The Raman-excited optical phonon, seen as periodic oscillation, is more pronounced than in \ce{Bi_2Se_3} and slightly red-shifted due to the heavier Te atoms (Supplementary information Fig. S10). After a sharp increase in reflectivity, the non-oscillatory part decreases within 2 ps via electron-phonon scattering and then monotonously increases again reaching a plateau at a pump-probe delay of 15 ps.  At longer delays, the reflectivity slowly decreases with a decay timescale of about 600 ps (shown in the Supplementary information). Such dynamics are very similar to those observed for 3 eV pump excitation \cite{Wang2016} that was attributed to the interplay between free carrier absorption and the band filling effect. 
\\
 
In comparison with \ce{Bi_2Te_3} and \ce{Bi_2Se_3}, the carrier dynamics in \ce{BSTS} (Fig.~\ref{fig:TI}c) show a much faster decay after the ultrafast build-up caused by THz excitation. The slow response observed in the other samples immediately after pump excitation is almost negligible. The relaxation time in \ce{BSTS} is short -- a few hundred fs -- which is very different compared to \ce{Bi_2Se_3}, \ce{Bi_2Te_3} and also with respect to the dynamics in \ce{BSTS} observed in studies employing above-bandgap excitation \cite{Onishi2015, Choi2018}. In all these cases, a pronounced relaxation component with a timescale of a few picoseconds is present. We ascribe the observed fast dynamics to the isolated response of Dirac fermions in the surface states of our TI system. We suggest that these fast surface-state relaxation dynamics we unveiled are the result of efficient coupling of carriers to phonons \cite{Hatch2011, Hsieh2011} and ultrafast surface-to-bulk scattering \cite{Wang2012}, and could hence be a general property of this material class. We note that the observed decay is significantly faster than the decay of surface-state carriers reported at 5~K, which was $\sim$1.5 ps \cite{Luo2019}. 
\\

To verify independently that the ultrafast relaxation in \ce{BSTS} originates from intraband dynamics within the topological SS, we have performed degenerate optical pump-probe measurements using pump pulses with a photon energy of 1.5 eV (Fig.~\ref{fig:TI}c). This scheme leads to the simultaneous excitation of bulk and surface states in all samples. In the case of \ce{Bi_2Te_3} and \ce{Bi_2Se_3} (Fig.~\ref{fig:TI}a-b), we see almost identical relaxation dynamics for optical and THz excitation, confirming that the dominant contribution to the dynamics originates from bulk states in bulk-metallic TIs. For BSTS we, in contrast to the below band-gap THz excitation, observe also a much slower dynamic on picosecond time scales in agreement with earlier observations for above band-gap excitation  \cite{Cheng2014}. The significantly faster decay after photoexcitation with below-bandgap THz photons, in comparison with excitation with above-bandgap photons, clearly suggests that the surface-state Dirac fermions have a much shorter lifetime. 
\\

\begin{figure}[h]
\includegraphics{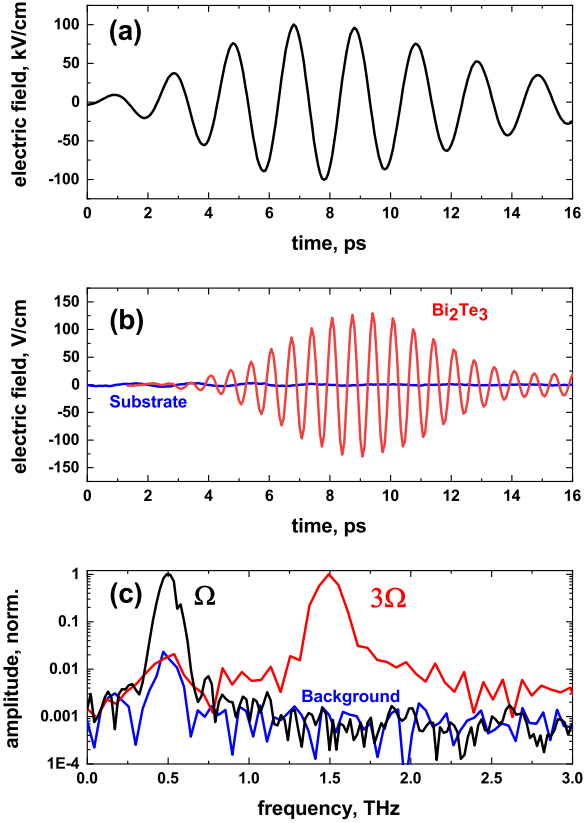}
\caption{\label{fig:TDS} Time-domain waveforms of the fundamental radiation electric field (a) and the third harmonic field generated in \ce{Bi_2Te_3}, and the \ce{Al_2O_3} substrate for reference (b). Amplitude spectra of the fundamental radiation and THG in \ce{Bi_2Te_3} and substrate (c).}
\end{figure}
 

In order to obtain additional insights into the relaxation dynamics of Dirac fermions in TI surface states, we study THz high-harmonic generation (HHG). Strong THz high-harmonic generation (HHG) has been observed in a number of Dirac materials: in graphene \cite{Hassan2018,Hassan2020,Deinert2021}, in the 3D Dirac semimetal \ce{Cd_3As_2} \cite{Kovalev2019, Cheng2019}, and in the prototypical TI \ce{Bi_2Se_3}, where it was shown that the THz harmonics originate from the Dirac electrons at the surface \cite{Giorgianni2016}. The underlying mechanism for THz HHG by Dirac fermions can be described by a thermodynamic nonlinearity mechanism that relies crucially on the ultrafast modulation of the THz absorption. This mechanism is enabled by efficient heating and subsequent cooling of the electronic system on fs to ps timescales upon interaction with strong THz fields \cite{Mics2015, Hassan2018, Hassan2020}. Therefore, the THz HHG behaviour is intricately related to the characteristic timescales of carrier dynamics of Dirac fermions. In particular, we point out that when cooling takes several picoseconds, this leads to a reduction of the nonlinearity coefficient and strong saturation effects for increasing incident field strengths, as observed recently for graphene under strong THz excitation \cite{Deinert2021} (see also Supplementary information). 
\\

In Fig.~\ref{fig:TDS}, we show measurements of fundamental and third-harmonic fields measured by electro-optic sampling for \ce{Bi_2Te_3}, together with the respective amplitude spectra. The details of the THz HHG experimental setup are show in the Supplementary information. We observe a clear third-harmonic signal from \ce{Bi_2Te_3} at 1.5 THz (Fig.~\ref{fig:TDS}b) upon driving the sample with the 0.5 THz fundamental with a peak field strength of 100 kV/cm. 
In Fig.~\ref{fig:TDS}c the resulting amplitude spectra of the fundamental beam and the THG in both TI and substrate are shown. In these experiments the THG polarization is collinear with the polarization of the linear fundamental beam polarization. The third-harmonic field strength scales down linearly with fundamental beam ellipticity reaching zero in case of circularly polarized pump (see the Supplementary information). 
\\


\begin{figure}[h]
\includegraphics{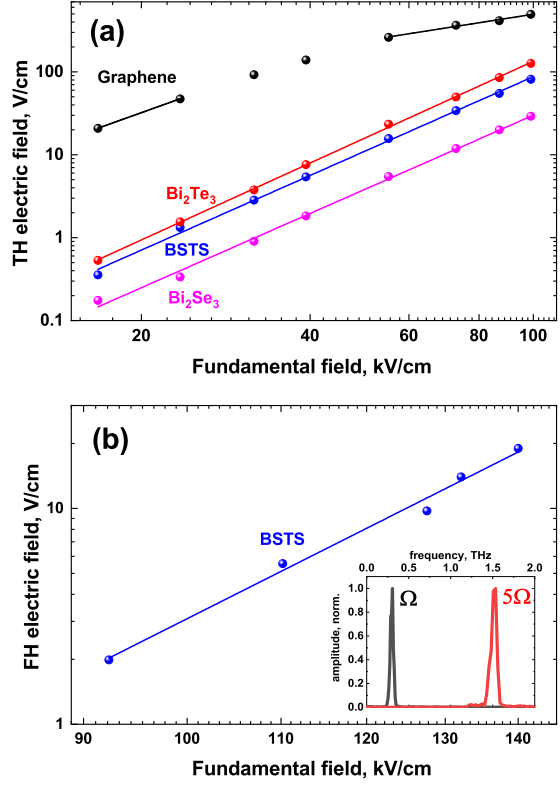}
\caption{\label{fig:Comparison} (a) The peak values of third harmonic fields generated in topological insulators and graphene obtained for different driving fields of fundamental radiation at 500 GHz central frequency (dots). Solid lines are linear fits providing the scaling law: $1.06\pm 0.08$  for graphene at highest field and 2.3 at lowest fields, $3.08\pm 0.03$ for \ce{Bi_2Te_3}, $2.99\pm 0.06$ for \ce{BSTS}, and $2.98\pm 0.07$ for \ce{Bi_2Se_3}; (b) fields of fifth harmonic generated in \ce{BSTS}, the slope of the linear fit is $5.2 \pm 0.3$; The inset shows amplitude spectra of fundamental and fifth harmonic.}
\end{figure}

In Fig.~\ref{fig:Comparison}a, we show the peak field strengths of the third-harmonic signal generated in \ce{Bi_2Se_3}, \ce{Bi_2Te_3} and \ce{BSTS} as a function of the field strength of the fundamental radiation. For comparison we show THz third-harmonic signal measurements from $p$-doped graphene with around $10^{13}$ cm$^{-2}$ charge carrier concentration. For fundamental field strengths between 60 and 100 kV/cm the third-harmonic signal from graphene exhibits clear saturation behavior and begins to scale linearly with the fundamental field, rather than the expected cubic scaling for third-harmonic generation in the perturbative regime. Even in the lowest range of driving fields, between 15 and 25 kV/cm, third-harmonic generation in graphene shows a power law scaling with an exponent of 2.3, indicating that the perturbative regime is only maintained at driving fields below 10 kV/cm. These observations for graphene are in agreement with earlier publications \cite{Hassan2018, Cheng2019, Deinert2021}. The strong saturation at the highest field strengths has been partially ascribed to a slowing down of the cooling of hot Dirac carriers, decreasing the nonlinear conversion efficiency \cite{Deinert2021}.\\

Compared to graphene, the TI samples show fundamentally different behaviour: all topological insulators demonstrate a clear cubic dependence of the third-harmonic field across the whole accessible field regime meaning that THz-induced nonlinear processes are far away from the saturation regime. In order to verify this perturbative behavior further, we show in Fig.~\ref{fig:Comparison}b the field strength of the fifth harmonic as a function of the fundamental field strength for the \ce{BSTS} sample. Using a power law fit, the fifth-harmonic fields scale with an exponent of $5.2 \pm 0.3$ as a function the fundamental field, which is clear evidence that the FHG process is also in the perturbative regime. This purely perturbative behavior all the way up to incident field strengths of 140 kV/cm can be ascribed (partially) to the sub-picosecond cooling dynamics we observed in our TPOP experiments: the ultrafast surface-state relaxation of the TI samples prevents heat accumulation effects within the comparably long duration (several picoseconds) of the THz excitation pulse.
An additional reason for the extended perturbative regime for the tested TIs could be the lower nonlinearity, which is likely due to the lower Fermi velocity compared with graphene \cite{Tang2013}. This allows for driving the TIs samples with higher fields before saturation is observed.
\\

The observation of perturbative HHG in TIs is highly interesting, because it could mean that a higher conversion efficiency can be obtained for increased incident field strengths. We observe that the highest overall conversion efficiency for graphene is about 0.5 $\%$ in field, which is consistent with other studies \cite{Hassan2018, Deinert2021}. However, saturation effects at high fields may  impose an efficiency limit. For the TI samples, we observe maximum third-order conversion efficiencies of  0.13, 0.08, and 0.03 $\%$ in field for \ce{Bi_2Te_3}, \ce{BSTS}, and \ce{Bi_2Se_3}, respectively. A maximum conversion efficiency for fifth harmonic generation (FHG) of around 0.014 $\%$ at 140 kV/cm fundamental field strength is observed. These efficiencies are lower than the one obtained for graphene. However, the purely perturbative scaling indicates that higher conversion efficiencies than for graphene are achievable by further increasing the incident field, which is of great technological interest. 
\\

 
In conclusion, we experimentally isolated the dynamical response of Dirac fermions in surface states of TIs that are excited by THz light, and observe ultrafast relaxation dynamics on a timescale of a few hundred femtoseconds. This decay is significantly faster than the dynamics of photoexcited carriers in bulk states, and likely originates from efficient phonon-assisted scattering either via surface intraband or surface-bulk interband transitions. In agreement with such fast relaxation, we observe no saturation effects in THz harmonic generation up to the highest field of 140 kV/cm, where the graphene harmonic response is already strongly saturated. Our findings are of high technological relevance since they indicate that the nonlinear conversion efficiencies in the investigated TI's can, in contrast to graphene, potentially be scaled to unprecedented values e.g. by employing metamaterials to locally enhance the electric fields \cite{Deinert2021}. Furthermore, the observed fast relaxation of surface-state Dirac fermions could open up new opportunities for ultrafast spintronic devices based on topological insulators.
\\

 \section{Acknowledgements}
 The authors are grateful to A.V. Shepelev and S.A. Tarasenko for useful discussions.
 Parts of this research were carried out at ELBE at the Helmholtz-Zentrum Dresden-Rossendorf e.V., a member of the Helmholtz Association. The films are grown in IRE RAS within the framework of the state task. This work was supported by the RFBR grants Nos. 18-29-20101, 19-02-00598.
 N.A., S.K., and I.I. acknowledge support from the European Union’s Horizon 2020 research and innovation program under grant agreement No. 737038 (TRANSPIRE).
 T.V.A.G.O. and L.M.E. acknowledge the support by the W\"urzburg-Dresden Cluster of Excellence on Complexity and Topology in Quantum Matter (ct.qmat).
 K.-J.T. acknowledges funding from the European Union's Horizon 2020 research and innovation program under Grant Agreement No. 804349 (ERC StG CUHL) and financial support through the MAINZ Visiting Professorship. ICN2 was supported by the Severo Ochoa program from Spanish MINECO Grant No. SEV-2017-0706.


\section{Data availability}
The data that support the findings of this study are available from the corresponding authors upon reasonable request.

\bibliography{Reference.bib}

\section{Competing Interests}
The Authors declare no Competing Financial or Non-Financial Interests.

\section{Author contributions}
The study was performed under the supervision of S.K. and K.A.K.
P.I.K. fabricated the samples. 
K.A.K, I.I., D.A.S. and G.Kh.K characterized the samples with linear THz transmission spectroscopy.
T.V.A.G.dO. and L.M.E. performed topography characterization.
S.K., K.J.T., H.H., D.T. and M.G. discussed and interpreted the experimental results.
S.K. and I.I. performed  pump-probe measurements.
S.K., I.I., A.P., J.C.D., N.A. M.C., M.B. performed THz HHG measurements.
S.K., M.G. and K.J.T. wrote the manuscript with contributions from J.C.D., I.I., A.P., T.V.A.G.dO., L.M.E., K.A.K., G.Kh.K., H.H., and D.T. 
All authors discussed the results and commented on the manuscript.

\end{document}